\def\lae{\mathrel{<\kern-1.0em\lower0.9ex\hbox{$\sim$}}}
\def\gae{\mathrel{>\kern-1.0em\lower0.9ex\hbox{$\sim$}}}

\newcommand{\be}{\begin{equation}}
\newcommand{\ee}{\end{equation}}

\documentclass[apj]{emulateapj}
\bibpunct[; ]{(}{)}{;}{a}{}{;}
\usepackage{apjfonts}

\slugcomment{Accepted for publication in  ApJ Letters}

\shortauthors{JORD\'AN ET AL.}
\shorttitle{GCLF TRENDS}

\begin{document}
 
\title{Trends in the Globular Cluster Luminosity Function of Early-Type
       Galaxies\altaffilmark{1}}

\author{
Andr\'es Jord\'an\altaffilmark{2},
Dean E. McLaughlin\altaffilmark{3},
Patrick C\^ot\'e\altaffilmark{4},
Laura Ferrarese\altaffilmark{4},
Eric W. Peng\altaffilmark{4},
John P. Blakeslee\altaffilmark{5},
Simona Mei\altaffilmark{6},
Daniela Villegas\altaffilmark{2,7},
David Merritt\altaffilmark{8},
John L. Tonry\altaffilmark{9},
Michael J. West\altaffilmark{10,11}
}

\begin{abstract}
We present results from a study of the globular cluster luminosity
function (GCLF) in a sample of 89 early-type galaxies observed as part
of the ACS Virgo Cluster Survey. Using a Gaussian parametrization of
the GCLF, we find a highly significant correlation between the GCLF
dispersion, $\sigma$, and the galaxy luminosity, $M_{B,{\rm gal}}$, in
the sense that the GC systems in fainter galaxies have narrower
luminosity functions. The GCLF dispersions in the Milky Way and M31 are 
fully consistent with this trend, implying that the correlation between sigma 
and galaxy luminosity is more fundamental than older suggestions that GCLF 
shape is a function of galaxy Hubble type.
We show that the $\sigma - M_{B,{\rm gal}}$ relation 
results from a bonafide narrowing of the
distribution of (logarithmic) cluster masses in fainter
galaxies.
We further show that this behavior is mirrored by a
steepening of the GC mass function for relatively high masses,
${\cal M} \ga 3\times10^5\,{\cal M}_\odot$, a mass regime in which
the shape of the
GCLF is not strongly affected by dynamical evolution over a Hubble time. We 
argue that this trend arises from variations in initial conditions
and requires explanation by theories of cluster formation.
Finally, we confirm that in bright galaxies, the GCLF ``turns over" at the
canonical mass scale of
${\cal M}_{\rm TO}\simeq 2\times 10^5\,{\cal M}_\odot$. However, we find 
that ${\cal M}_{\rm TO}$ scatters to lower values
($\approx$ 1-2$\times 10^5\,{\cal M}_\odot$) in galaxies fainter than
$M_{B,{\rm gal}}\ga -18.5$, an important consideration if 
the GCLF is to be used as a distance indicator for dwarf ellipticals.
\end{abstract}


\altaffiltext{1}{Based on observations with the NASA/ESA
{\it Hubble Space Telescope}
obtained at the Space Telescope Science Institute, which is operated
by the Association
of Universities for Research in Astronomy, Inc., under
NASA contract NAS 5-26555}
\altaffiltext{2}{European Southern Observatory,
Karl-Schwarzschild-Stra{\ss}e 2, 85748 Garching bei M\"unchen, Germany;
ajordan@eso.org}
\altaffiltext{3}{Department of Physics \& Astronomy, University
of Leicester, Leicester, LE1 7RH, UK; dean.mclaughlin@astro.le.ac.uk}
\altaffiltext{4}{Herzberg Institute of Astrophysics, Victoria, 
BC V9E 2E7, Canada}
\altaffiltext{5}{Department of Physics and Astronomy,
Washington State University, 1245 Webster Hall, Pullman, WA 99163-2814}
\altaffiltext{6}{Department of Physics and Astronomy,
Johns Hopkins University, Baltimore, MD 21218}
\altaffiltext{7}{Departamento de Astronom\'{\i}a y Astrof\'{\i}sica, 
Pontificia Universidad Cat\'olica de Chile, 
Avenida Vicu\~na Mackenna 4860, Casilla 306, Santiago 22, Chile}
\altaffiltext{8}{Department of Physics, Rochester Institute of Technology,
84 Lomb Memorial Drive, Rochester, NY 14623}
\altaffiltext{9}{Institute of Astronomy, University of Hawaii,
2680 Woodlawn Drive, Honolulu, HI 96822}
\altaffiltext{10}{Department of Physics \& Astronomy, University of Hawaii,
Hilo, HI 96720}
\altaffiltext{11}{Gemini Observatory, Casilla 603, La Serena, Chile}

\section{Introduction}
\label{sec:intro}

The luminosity function of globular clusters (GCs) represents
one of the most remarkable features of these stellar systems.
The distribution of GC magnitudes, commonly referred to as the
GC luminosity function (GCLF), shows a turnover, or peak, at 
$M_V \simeq -7.5$ mag, corresponding to a mass of 
${\cal M} \simeq 2 \times 10^5 {\cal M}_\odot$. Observations have shown 
that this turnover is nearly invariant across and within galaxies, prompting
its widespread use as a distance indicator (see, e.g., Harris 2001).
Accounting for this nearly universal mass scale remains
an open problem for theories of GC formation and evolution.
It follows that establishing whether or not the GCLF as a whole is
universal --- i.e., whether its overall form depends on
host galaxy properties --- can help guide and
constrain theories for the formation and evolution
of galaxies and GC systems.

In this {\it Letter}, we present results from a study of 
the GCLFs of 89 early-type galaxies observed by HST as
part of the ACS Virgo Cluster Survey (ACSVCS; C\^ot\'e et al.~2004).
We find the clearest evidence to date for a correlation
between the width (i.e., Gaussian dispersion) of the GCLF and the
luminosity of the host galaxy; 
we also show that there is some downward scatter in the
{\it mass} scale of the GCLF turnover in galaxies fainter
than $M_{B,{\rm gal}}\ga -18.5$.
Focusing on the observed steepening of the GCLF
at the bright (high-mass) end in the faint galaxies,
we argue that this behavior was probably imprinted at the time of GC
formation. A more detailed discussion of the whole GCLF, 
including the faint (low-mass) end and the
role that long-term dynamical evolution plays in that regime, is deferred to a
subsequent paper (Jord\'an et~al.~2006, hereafter J06). That paper
presents our data in full and gives details of our analysis techniques,
including modeling of the GCLFs with a new, non-Gaussian, physically
motivated fitting function.

\section{Observations and Analysis}
\label{sec:data}

One hundred early-type galaxies in the Virgo cluster 
were observed in the ACSVCS (C\^ot\'e et~al. 2004).
Each galaxy was imaged for 750 s in the F475W bandpass
($\simeq$ Sloan~$g$) and for 1210 s in F850LP ($\simeq$ Sloan~$z$).
Reductions were performed as described in
Jord\'an et~al. (2004). In what follows, we use $g$ and $z$ as
shorthand to refer to the F475W and F850LP filters.

One of the main scientific objectives of the ACSVCS is the study of
GC systems, and thus we have developed methods
to: (1) discard foreground stars and background galaxies from the
totality of observed sources around each target galaxy in the survey;
and (2) estimate the
level of residual fore- and background contamination in the remaining sources
designated as candidate GCs. These procedures are described and illustrated
by Peng et al.~(2006a; their \S2.2 and Figure~1), and discussed in detail in
the GCLF context in J06. In the latter paper, we also
examine the effects of using alternate selection
criteria to define GC samples, and show that the results
presented here are fully robust against such subtleties.

Of the 100 galaxies in the ACSVCS, we restrict our analysis to those that
have more than five GCs, as estimated by subtracting the total number
of expected contaminants from the full list of GC candidates for each galaxy.
We additionally eliminate two galaxies for which we were
unable to obtain useful measurements of the GCLF parameters. This leaves
a final sample of 89 galaxies which are studied here, and in J06.

Also as part of the ACSVCS, we have measured the distances to 84
of our target galaxies using the method of surface
brightness fluctuations (SBF; Mei et~al. 2006). We use these SBF distances to
transform the observed
GC and galaxy magnitudes into absolute ones whenever possible. For those
galaxies lacking an SBF distance, 
we adopt the mean distance modulus to the
Virgo cluster: $\langle (m-M)_0 \rangle = 31.09$~mag, or
$\langle D \rangle=16.5$~Mpc (see Mei et~al. 2005, 2006).

We use an approach similar to that of Secker \& Harris (1993) to
characterize the GCLFs: parametric models are fitted to the observed
luminosity functions via a maximum-likelihood method that takes into account
photometric errors, incompleteness, and the luminosity function of 
contaminants. Full technical details are given in J06, where we consider
two parametric models for the GCLF. The first, on which this paper will focus,
is the standard Gaussian distribution,
\begin{equation}
dN/dz = N_{\rm tot} \, \big( 2\pi\,\sigma_z^2 \big)^{-1/2} \
            \exp \big[- (z-\mu_z)^2/2\sigma_z^2 \big]\ .
\label{eq:gauss}
\end{equation}
The second is a simple analytical modification of a Schechter (1976) function
designed to account for the effects of cluster evaporation (two-body
relaxation) on a GC mass function that is assumed to have initially resembled
that of the young clusters forming today in local mergers and starbursts. Full
details on these two models are given in J06, where we fit each of them
to the separate $g$- and $z$-band GCLFs of our 89 program galaxies. In this
{\it Letter} we present only the results of Gaussian fits to the $z$-band
GCLFs.

\section{Results}
\label{sec:results}

\begin{figure}
\epsscale{1.0}
\plotone{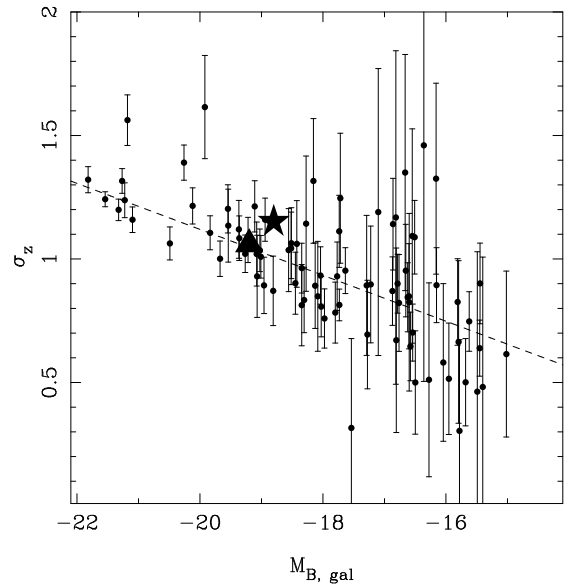}
\caption{Gaussian dispersion, $\sigma_z$, versus galaxy, $M_{B,{\rm gal}}$,
for the $z$-band GCLFs of 89 ACSVCS galaxies. The GCLF width varies
systematically, being narrower in fainter galaxies. The two anomalously high
points at $M_{B, {\rm gal}}=-21.2$ and $-19.9$ correspond to the galaxies
VCC 798 and VCC 2095, both of which have large excesses of faint, diffuse
clusters (Peng et al.~2006b). The large star is plotted at the 
spheroid luminosity (de Vaucouleurs \& Pence 1978) and GCLF dispersion (Harris
2001) of the Milky Way. The large triangle marks the bulge luminosity (Kent
1989) and GCLF dispersion (Harris 2001) of M31.
\label{fig:sigma_B}}
\end{figure}

Figure~\ref{fig:sigma_B} shows our main result: GCLFs are narrower in
lower-luminosity galaxies. The straight line in this plot of Gaussian
dispersion against absolute galaxy magnitude 
shows the least-squares fit
\begin{equation}
\sigma_z = (1.12\pm0.01) - (0.093 \pm 0.006)(M_{B, {\rm gal}}+20) \ .
\label{eq:sigmaz} 
\end{equation}

It has been reported before that the GCLFs in
lower-luminosity galaxies tend to show somewhat lower dispersions
(e.g., Kundu \& Whitmore 2001). However, the size and homogeneity of
the ACSVCS dataset make this
the most convincing demonstration to date of a
continuous trend in GCLF shape over a range of $\ga\! 400$ in galaxy
luminosity. Monte Carlo simulations and alternate constructions of GCLF
samples show that the observed decrease in dispersion is {\it not} 
an artifact of small-number statistics in the faint galaxies (J06).

Past investigations have pointed to a dependence of the GCLF dispersion on
the Hubble type of the GC host galaxies (e.g., Harris
1991). Figure~\ref{fig:sigma_B}
includes datapoints at the location of the bulge magnitude and GCLF
dispersion of the Milky Way (large star) and M31 (large triangle). 
Since both systems fall comfortably on the relation defined by
our data for early-type galaxies, we conclude that the underlying
fundamental correlation is one between $\sigma$ and $M_{B,\rm gal}$,
rather than between $\sigma$ and Hubble type.

A natural question at this point is whether the observed trend in GCLF 
dispersion with galaxy magnitude implies a similar trend in the GC
{\it mass} function. This is not a foregone conclusion, for the following
reason. GC systems are known to have systematically redder
and broader (or more strongly bimodal) color distributions in brighter
galaxies than in faint ones (see, e.g., Peng et al.~2006a). Equivalently, 
GCs in giant galaxies are more metal-rich on average, and have larger
dispersions in [Fe/H], than those in low-mass dwarfs.
Since cluster
mass-to-light ratios, $\Upsilon$, are functions of [Fe/H] in
general, it is conceivable that the average GC $\Upsilon$ could
change systematically in going from bright galaxies to fainter ones,
and that the spread of $\Upsilon$ values within a single GC system     
could also vary systematically as a function of galaxy magnitude.
The possibility then exists
that narrower GCLFs for faint galaxies might result from these systematics
in $\Upsilon$ combined with a more nearly invariant spread in GC
masses.
We can show easily, however, that this is not the case.

The systematics in $\Upsilon$ vs.~[Fe/H] just mentioned are
also a function of wavelength. In bluer filters, such as $B$, $V$, or $g$,
mass-to-light ratios of old stellar systems 
do change significantly (increasing by factors 
of two or more) in going from cluster metallicities ${\rm [Fe/H]} \la -2$ to
${\rm [Fe/H]}=0$, typical of GCs. But at the much redder wavelengths of our
$z$-band data ($\lambda_{\rm pivot} \simeq 9055$~\AA; Sirianni et~al. 2005), 
this strong metallicity dependence almost completely disappears. We have used
the PEGASE population-synthesis 
model of Fioc \& Rocca-Volmerange (1997) to compute $\Upsilon_z$ as a function
of metallicity for clusters with a Kennicutt (1983)
stellar IMF and various fixed ages $\tau$. For $\tau=13$~Gyr, we find
that $\Upsilon_z\simeq 1.6\ {\cal M}_\odot\,L_\odot^{-1}$ at an extreme
${\rm [Fe/H]}=-2.3$, decreasing to a minimum of
$\Upsilon_z\simeq 1.5\ {\cal M}_\odot\,L_\odot^{-1}$ at
${\rm [Fe/H]}\simeq-0.7$, and then increasing slightly to
$\Upsilon_z=1.7\ {\cal M}_\odot\,L_\odot^{-1}$ at
${\rm [Fe/H]}=0$. In other words, we always have
$\Upsilon_z\approx 1.6\pm0.1$ for any of the globular clusters
in any of our sample galaxies --- no matter how red or blue the clusters are,
or how broad or narrow the GC color/metallicity distribution.
Comparably small ranges of $\Upsilon_z$ result if younger GC ages or different
reasonable stellar IMFs are assumed.

The effect of variations in mass-to-light ratio on the width of the GCLF at
NIR wavelengths is therefore completely negligible. From
Figure~\ref{fig:sigma_B}, we have that $\sigma(\log\,L_z)=\sigma_z/2.5\ga0.2$
in our galaxies, whereas the discussion above implies that
$\Upsilon_z({\rm max})/\Upsilon_z({\rm min})\sim 1.13$ for our GCs, such that
the dispersion of mass-to-light ratios in any one system is always
$\sigma(\log\,\Upsilon_z)<0.055$ at an absolute maximum.
The intrinsic dispersion of logarithmic GC {\it masses},
$\sigma(\log\,{\cal M}) =
    [ \sigma^2(\log\,L_z) - \sigma^2(\log\,\Upsilon_z) ]^{1/2}$,
is thus never more than $\sim\!4\%$ different from the
observed $\sigma(\log\,L_z)$. 
We conclude, unavoidably, that the steady decrease of $\sigma_z$ by more than
50\% from the brightest giants to the faintest dwarfs in
Figure~\ref{fig:sigma_B} is an accurate reflection of
just such a trend in the intrinsic GC mass distributions.

\begin{figure}[!t]
\epsscale{1.0}
\plotone{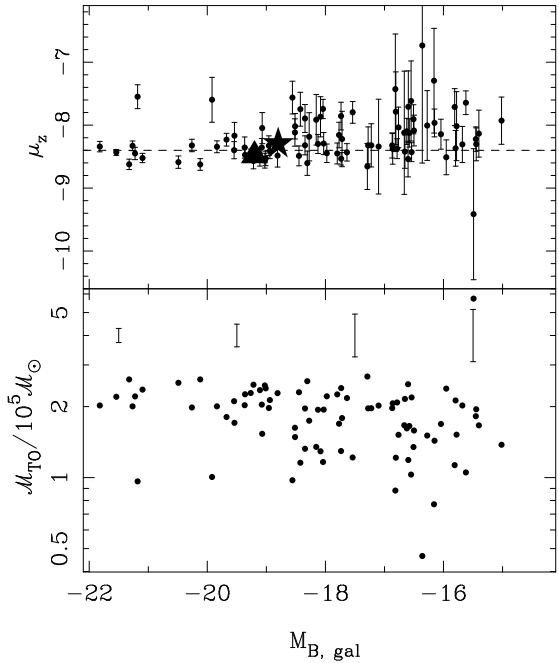}
\caption{
({\it Top}) GCLF turnover magnitude (absolute mean $\mu_z$) versus galaxy
magnitude, $M_{B,{\rm gal}}$, from Gaussian fits to 89 $z$-band GCLFs in the
ACSVCS. The outlying points at $M_{B,{\rm gal}}=-21.2$ and $-19.9$ are
VCC 798 and VCC 2095, which have large excesses of faint, diffuse clusters
(Peng et al.~2006b). The star and triangle show $\mu_z$ values for the Milky
Way and M31, estimated from their $V$-band peaks (Harris 2001) by applying an
average $(V-z)$ color estimated from the PEGASE population-synthesis code
(Fioc \& Rocca-Volmerange 1997). 
({\it Bottom}) Turnover {\it mass} ${\cal M}_{\rm TO}$ corresponding to the
fitted $\mu_z$, obtained by applying an average GC $\Upsilon_z$ computed for
each galaxy using the PEGASE model. Typical errorbars on ${\cal M}_{\rm TO}$
as a function of galaxy magnitude are indicated. 
\label{fig:mu_B}}
\end{figure}

We now turn our attention to the GCLF turnover magnitude.
The upper panel of Figure~\ref{fig:mu_B} shows the mean GC absolute
magnitude $\mu_z$ from the Gaussian fits to our GCLFs, versus
host galaxy $M_{B,{\rm gal}}$. 
The horizontal line in this plot is drawn at a level typical of
galaxies brighter than $M_{B,{\rm gal}}\la -18.5$: $\mu_z=-8.4$. Given
a typical $\Upsilon_z\simeq1.5\ {\cal M}_\odot\,L_\odot^{-1}$ in these
galaxies (for GC ages 13 Gyr and an average
${\rm [Fe/H]}\approx-1$), this
corresponds to a cluster mass scale of ${\cal M}_{\rm TO}\simeq 2.2\times10^5\
{\cal M}_\odot$. 
Estimates of the $z$-band GCLF turnovers in the Milky Way and M31
are shown by the large star and large triangle, as in Figure~\ref{fig:sigma_B}.
In the lower panel of Figure \ref{fig:mu_B} we plot the turnover
masses ${\cal M}_{\rm TO}$ obtained from the fitted $\mu_z$ using 
PEGASE model mass-to-light ratios.
As we have discussed, $z$-band luminosities are very good proxies for total
GC masses, so this graph is essentially a mirror image of the one above it.

Figure~\ref{fig:mu_B} shows that there is no strong or systematic 
variation in $\mu_z$ or
${\cal M}_{\rm TO}$ to match that seen for $\sigma_z$ 
(Figure~\ref{fig:sigma_B}). Nevertheless,
there is a clear tendency for the GCLF turnovers of 
galaxies fainter than $M_{B,{\rm gal}}\ga -18.5$ to scatter to somewhat
fainter (less massive) values than is typical of the bright giants.
The difference in mass is a factor of $\approx\!1.5$ on average,
but it ranges apparently randomly, from a factor of 1 (i.e., no
difference) up to factors slightly greater than 2 in some cases.
Note that there is a healthy mix of E and S0 or dE and dS0 galaxies
at all magnitudes in our ACSVCS sample (see Table 1 of C\^ot\'e et al.~2004).
We find no tendency for any particular Hubble type to scatter
preferentially away from $\mu_z=-8.4$ or ${\cal M}_{\rm TO} = 
2.2 \times 10^5 M_\odot$ in Figure \ref{fig:mu_B}.

The lower mean value for  ${\cal M}_{\rm TO}$ at faint $M_{B,{\rm gal}}$
clearly can impact the use of the GCLF as a standard candle for dwarf
galaxies. On the other hand, the effect is wavelength-dependent. Publicly
available codes such as PEGASE can be used easily to show that in bluer
bandpasses such as $g$ (or the closely related $V$, which is more standard
for such studies), the slight decrease we find for the average GC
turnover {\it mass} in fainter galaxies is balanced by a comparable
decrease in the typical GC mass-to-light ratio (because of the lower
cluster metallicities), so that the mean turnover {\it magnitude} does not
vary as in the $z$ band. We have also confirmed this directly from our own
ACSVCS data. In J06, we obtain plots analogous to Figures \ref{fig:sigma_B}
and \ref{fig:mu_B} from fits to the $g$-band GCLFs of our galaxies. The
results fully support all of our conclusions here. It is
particularly worth noting that we find a relation identical to equation
(\ref{eq:sigmaz}) for the dependence of $g$-band GCLF dispersion on parent
galaxy luminosity.

\section{Discussion}
\label{sec:discussion}

An obvious question prompted by Figure \ref{fig:sigma_B} is whether the
correlation between $\sigma_z$ and $M_{B,{\rm gal}}$ was established
at the time of cluster formation or built up afterwards as GCLFs were
modified by the dynamical destruction of GCs over a Hubble
time. We favor the first interpretation.

Star clusters can be destroyed over Gyr timescales as a result of
mass loss driven by stellar evolution, dynamical friction, gravitational
shocks, and internal two-body
relaxation (evaporation) --- processes that have been studied in
detail by several 
groups. Recent discussions, centered specifically on how these affect
the GCLF, can be found in Fall \& Zhang (2001) and Vesperini (2000, 2001).
Fall \& Zhang in particular show that, while stellar evolution and
gravitational shocks certainly deplete the total number of GCs in a
galaxy, they do 
not significantly alter the overall shape of the GCLF. Evaporation, on the
other hand, {\it can} change the shape of the GCLF, but significantly so only
for cluster masses ${\cal M}\la 2$-$3\times10^5\,{\cal M}_\odot$, i.e.,
below the typical GCLF turnover mass. 

In the theoretical treatments of Fall \& Zhang, Vesperini, and many
others, the evaporation rate is independent of cluster mass, which ultimately
drives the {\it low-mass} side of the GCLF to a universal
shape: a simple exponential $dN/dz \propto 10^{-0.4\,z}$ for the
number of GCs per unit magnitude fainter than the turnover (equivalent to
a flat distribution for the number of GCs per unit linear luminosity or
mass). But fitting a Gaussian model to the GCLF, as we have done
here, tacitly assumes that the distribution is symmetric. The
results in Figure \ref{fig:sigma_B} might therefore seem to imply that
both the bright side {\it and} the faint side of the GCLF become progressively
steeper in fainter galaxies.
However, various observational uncertainties make it difficult to
determine precisely the form of the faintest tail of the GCLF. Thus, in J06
we show that good fits to our GCLFs can also be obtained using an alternate
model with a universal exponential shape at 
magnitudes fainter than the turnover --- and that the downward scatter in
${\cal M}_{\rm TO}$ for faint galaxies persists in such a model and so is
not an artifact of any assumed Gaussian symmetry. Here we concern
ourselves only with the brighter half of the GCLF, which is observationally
better defined.

We have performed maximum-likelihood fits of
exponential models $dN/dz \propto 10^{0.4 (\beta_z-1)\,z}$ (corresponding to
power-law {\it mass} distributions, $dN/d{\cal M}\propto {\cal M}^{-\beta_z}$)
to the GCLFs at absolute magnitudes $-8.7 \ga z \ga -10.8$
(cluster masses $\simeq\! 3\times10^5$--$2\times10^6\ {\cal M}_\odot$)
in 66 of our galaxies. Such distributions accurately describe the
bright sides of giant-galaxy GCLFs (Harris \& Pudritz 1994;
Larsen et al. 2001), and with $\beta_z\simeq2$ they also give good
matches to the mass functions of young star clusters in nearby mergers
and starbursts (Zhang \& Fall 1999).

\begin{figure}
\epsscale{1.0}
\plotone{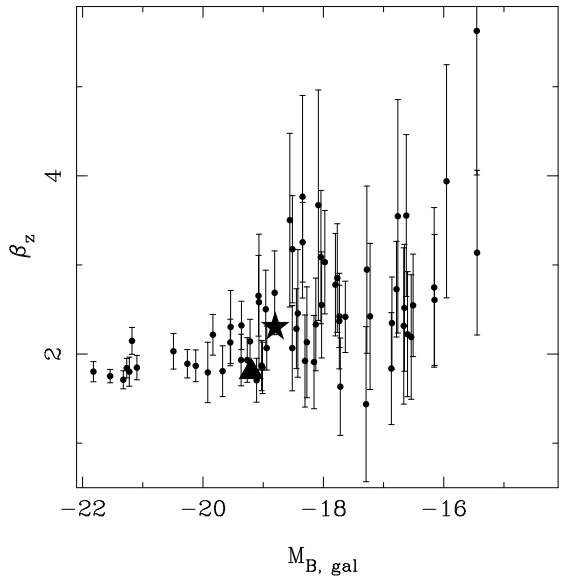}
\caption{Slope of the power law that best fits our $z$-band GCLF data,
$\beta_z$, for masses $3\times 10^5 \lae ({\cal M}/{\cal M}_\odot) \lae 2\times
10^6$, plotted against host galaxy absolute magnitude, $M_{B,{\rm gal}}$. 
The star and triangle show $\beta$ values for the Milky Way and M31 respectively, 
measured in the same mass regime using the data from Harris (1996) and Reed et~al. (1994)
assuming a $V$-band mass-to-light ratio $M/L_V = 2$.
The bright
side of the GCLF is steeper in fainter galaxies. 
\label{fig:power}}
\end{figure}

Figure~\ref{fig:power} shows the results of this exercise. 
There is a clear steepening in the 
power-law exponent, from $\beta_z\simeq 1.8$ in bright galaxies to 
$\beta_z\simeq3$ in the faintest systems. However the faint side of
the GCLF behaves in detail, the bright side alone suggests that smaller
galaxies were unable to form very massive clusters in the same {\it relative} 
proportions as giant galaxies.

A potential complication here is dynamical friction. A 
cluster of mass ${\cal M}$ on an orbit of radius $r$ in a galaxy with 
circular speed $V_c$ will spiral in to the center of the galaxy on a 
timescale $\tau_{\rm df} \propto {\cal M}^{-1} r^2 V_c$ (Binney \& Tremaine
1987). In the Milky Way and larger galaxies, $\tau_{\rm df} > 13$~Gyr
for all but the very most massive clusters at small radii, and thus 
dynamical friction does not significantly affect
their GCLFs (e.g., Fall \& Zhang 2001). In 
dwarfs with low $V_c$, however, $\tau_{\rm df}$ can be interestingly 
short for smaller GCs at larger $r$ --- suggesting, perhaps,
that the process might significantly deplete the bright side of the GCLF
in small galaxies and contribute to the type of
trend seen in Figure~\ref{fig:power}. However, Vesperini (2000, 2001) has
modeled the GCLF evolution over a Hubble time in galaxies with a wide range
of mass, and his results strongly suggest that dynamical
friction does {\it not} suffice to explain our observations. In
particular, the widths of the Gaussian GCLFs in his models do not decrease,
even in dwarf galaxies, to anywhere near the extent seen in the data.
Thus, any significant galaxy-to-galaxy variations in the shape of the GCLF
above the turnover mass probably reflect initial conditions
(see J06 for further discussion). 

In summary, the gradual narrowing of the GCLF as a function of galaxy
luminosity --- or the steepening of the 
mass distribution above the classic turnover point --- presents
a new constraint for theories of GC formation and evolution.
In our view, it is the cluster formation process in
particular that is likely to be most relevant to the observed behavior at 
the high-mass end of the GCLF. Exactly what factors might lead to more 
massive galaxies forming massive clusters in greater relative numbers,
is an open question of some interest. 

\acknowledgements

Support for program GO-9401 was provided 
through a grant from the Space
Telescope Science Institute, which is operated by the Association of
Universities for Research in Astronomy, Inc., under NASA contract NAS5-26555.

{\it Facility:} \facility{HST (ACS/WFC)}



\end{document}